\documentclass[journal]{IEEEtran}

\usepackage[utf8]{inputenc}
\usepackage{authblk}
\usepackage{graphicx}
\usepackage{caption}

\usepackage{url}
\usepackage{amsmath}
\usepackage{gensymb}
\title{Utilization of Additive Manufacturing for the Rapid Prototyping of C-Band RF Loads}
\author[1]{Garrett Mathesen}
\author[1]{Charlotte Wehner}
\author[1]{Julian Merrick}
\author[1]{Bradley Shirley}

\author[2]{Ronald Agustsson}
\author[2]{Robert Berry}
\author[2]{Amirari Diego}

\author[1]{Emilio A. Nanni}
\affil[1]{SLAC National Accelerator Laboratory, \textit{2575 Sand Hill Rd, Menlo Park, CA 94025}}
\affil[2]{RadiaBeam, \textit{21735 Stewart Street, Suite A, Santa Monica, CA 90404}}

\date{\today}

\begin{document}

\maketitle

\begin{abstract}
Additive manufacturing is a versatile technique that shows promise in providing quick and dynamic manufacturing for complex engineering problems. Research has been ongoing into the use of additive manufacturing for potential applications in radiofrequency (RF) component technologies. Here we present a method for developing an effective prototype load produced out of 316L stainless steel on a direct metal laser sintering machine. The model was tested within simulation software to verify the validity of the design. The load structure was manufactured utilizing an online digital manufacturing company, showing the viability of using easily accessible tools to manufacture RF structures. The produced load was able to produce an S$_{11}$ value of -22.8 dB at the C-band frequency of 5.712 GHz while under vacuum. In a high power test, the load was able to terminate a peak power of 8.1 MW. Discussion includes future applications of the present research and how it will help to improve the implementation of future accelerator concepts.
\end{abstract}

\section{Introduction}
\par Advances in additive manufacturing (AM) techniques have led to increased research efforts on its applicability to various engineering challenges and wider adoption within industry. Of particular interest is the ability of the use of AM technology to simplify the manufacturing process of typically complex parts. This can be seen through the heavy research being performed in the aerospace, automotive, and biomedical industries \cite{reviewOfAM}. Research is also being performed to develop engineered solutions for RF components, especially ones that require complex geometries such as in the development of terahertz RF structures \cite{terahertzAM} and X-band klystron and terminating load structures \cite{amKlystron,cern1}. 

\par
\begin{figure}[t]
\centering
\includegraphics[width=0.5\textwidth]{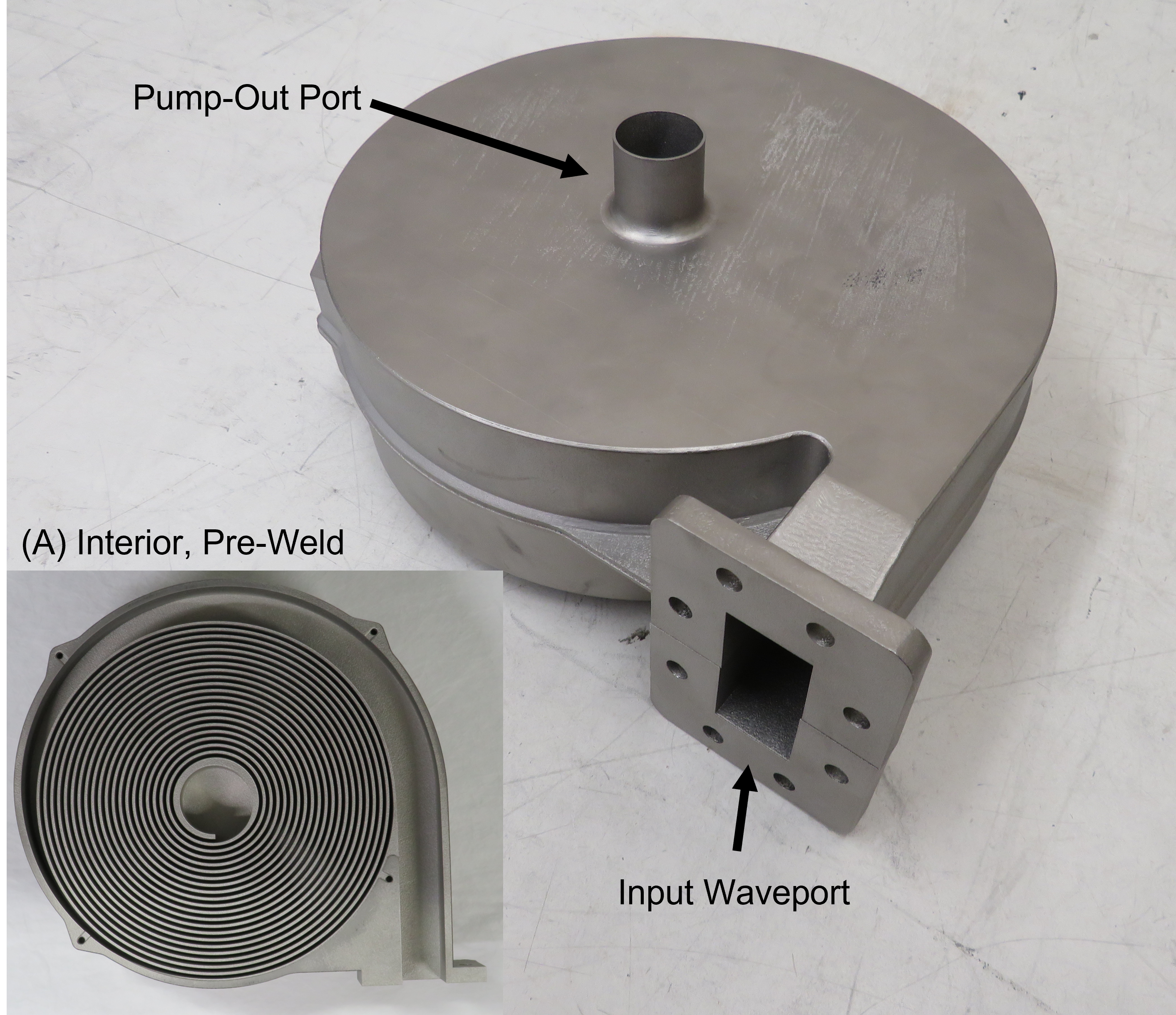}
\caption{Labelled image of printed spiral load, with interior structure shown in insert (A).}\label{fig:InstFlange}
\end{figure}

 The following paper continues this research by investigating the application of AM techniques to the creation of a C-band load. A background is given on how additive manufacturing works and what design constraints it adds. Motivation for the paper includes the potential future applications which will require the rapid development of complex high power loads. A methodology is outlined describing how the simulations and solid models were developed. Results of these simulations along with the results from tests of the realized spiral load (seen in Fig. \ref{fig:InstFlange}) are shown and discussed.

\begin{figure}[b]
\centering
\includegraphics[width=0.5\textwidth]{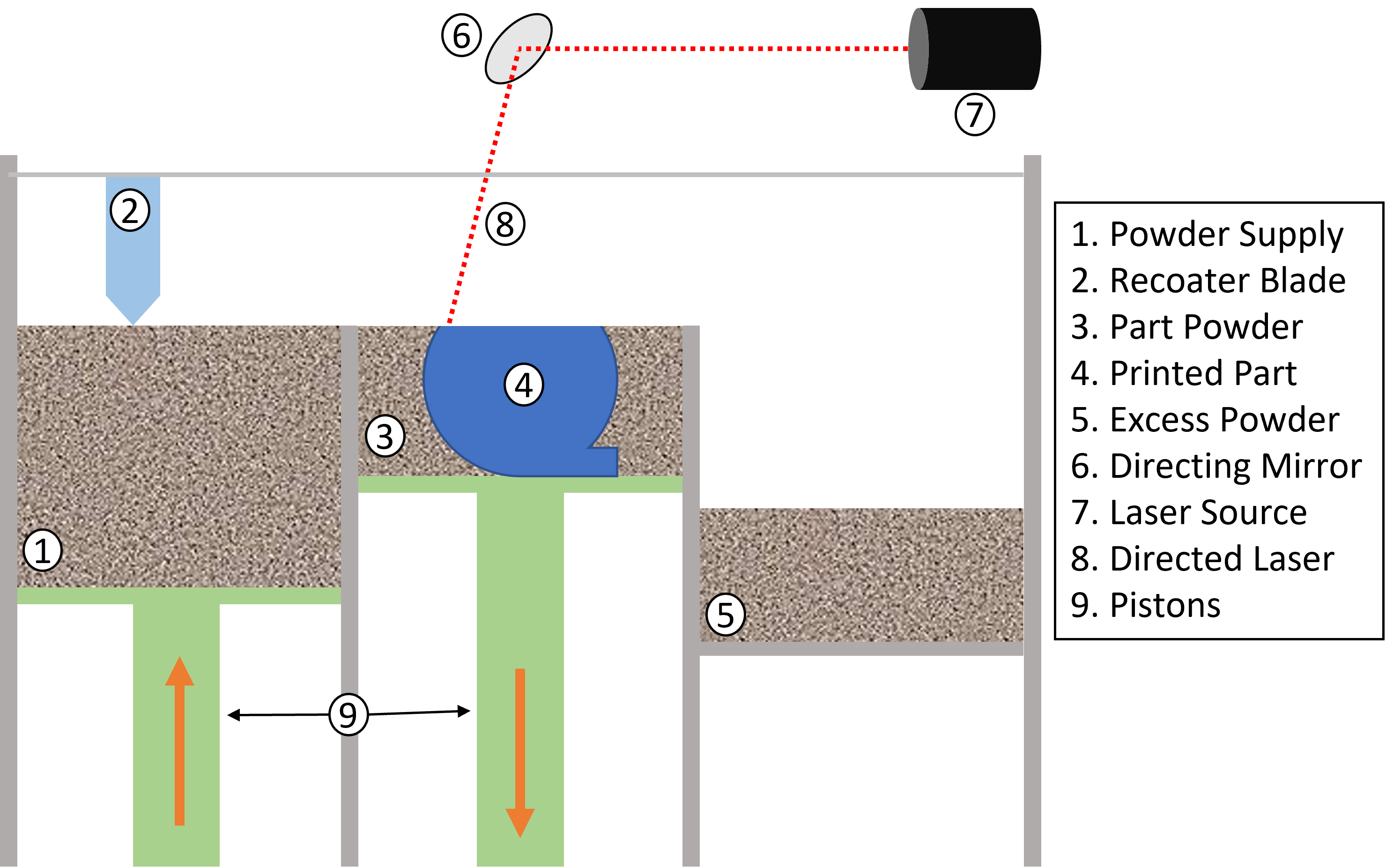}
\caption{Diagram showing the common components of a DMLS machine.}
\label{fig:DMLSDiagram}
\end{figure}

\subsection{Background}
\par The process of additive manufacturing can be simplified down to the core functionality of laying, binding, or solidifying layers of a given material on top of one another until a programmed design is produced. This process can be realized in various different mechanical methods, depending mainly on the material chosen. Common AM materials include polymers, ceramics, metals and composite materials. Within the realm of metal additive manufacturing (MAM) there are two main methods utilized for realizing MAM parts; direct energy deposit (DED) and powder bed fusion. DED utilizes a wire or powder feeder that allows for the stock to be melted by an energy source as the material is extruded. The energy source can be a laser, arc-welder, or electron beam depending on the implementation chosen. This method results in relatively cheap and quickly produced parts, but has the limitation of having low resolution and poor layer adhesion. The low resolution produced also inherently limits the complexity of the possible parts produced \cite{reviewOfAM}. 

\par Powder bed fusion, specifically a variety of the method known as direct metal laser sintering (DMLS), operates by utilizing a laser to sinter the metal rather than using a heat source to melt the metal particles together. The typical DMLS machine will operate by first raising the feeder bed so that a recoater blade can push the upper layer of the metal particles can coat the build platform. Any excess powder that does not cover the build plate will be pushed into an excess bin on the opposite side of the build plate from the feeder bin. A laser will then sinter the particles together in the desired pattern for the given layer. The build plate will then lower, another layer of particles will be coated on top of the last layer, and the process will repeat until the part is finalized. Fig. \ref{fig:DMLSDiagram} shows how these components are typically laid out. The lack of binding polymer necessary for other powder bed fusion methods also allows for higher resolution parts to be created \cite{reviewOfAM}. One of the major drawbacks of the DMLS method is that it can be high cost and the longer print times compared to other methods \cite{industrialAdoption}.

\par During the design phase, considerations were made with regard to the general limitations of current DMLS technologies and how AM processes would affect the printed design. A driving factor of the solid model design due to the AM process was the issue of internal support materials. Internal support material may be added in the event that an internal cavity has an excessive overhang due to the layer-by-layer nature of AM. Unlike external support material, internal supports are impractical if not impossible to remove once they are printed causing unexpected performance when compared with simulated performance.

\par While limitations with DMLS exist, these limitations are reflected in most other forms of additive manufacturing. DMLS also provides many advantages, as previously mentioned, that lends its use over others methods such as DED. Furthermore, theses identified drawbacks were mitigated through implementable design principles.

\subsection{Motivation}
\par Previous research has been performed in the use of AM to develop components for accelerator concepts. Of particular interest were interim reports from research performed by CERN demonstrating the way in which different materials and designs affected the performance of AM X-band load structures. This research showed promising results within the X-band range, with simulations showing the optimized waveguide geometry in Ti-6Al-4V performing -39.07 dB in the S$_{11}$ at a frequency of 11.9942 GHz \cite{cern1}. Several variations of the design were produced in different metals such as Ti-6Al-4V and 316L stainless steel. Of the eight models produced in preliminary findings, two of them were manufactured in 316L stainless steel and performed around -20 dB in the S$_{11}$ at 11.9942 GHz \cite{cern2}. These results showed the potential for further development in the utilization of this method for potential use within future collider and accelerator concepts. 

\par The Cool Copper Collider (C$^3$) is a proposed lepton-collider Higgs factory \cite{bai2021c,dasu2022strategy} which has a planned center C-band frequency of 5.712 GHz \cite{nanni2022c}. By using C-band over the more traditional S- and UHF-band seen in many high energy colliders currently, it greatly reduces the necessary size to obtain TeV-scale accelerators \cite{bane2018advanced}. By utilizing additive manufacturing in the production process, it will allow for rapid and flexible manufacturing of the large number of terminating loads that will be necessary for such a collider.

\par A third party, prototype manufacturing company was used for fabrication of the MAM parts of the load. This was done to demonstrate the practicality of using standard manufacturing pipelines for creation high power RF components. As DMLS machines are expensive to procure and require additional time and money for maintenance, using a third-party company with DMLS as part of their standard offerings greatly reduced the barrier to entry for the research performed.

\captionsetup{belowskip=0pt,aboveskip=0pt}

\captionsetup{labelsep=period}

\section{Methodology}

\par The design for the load structure was based on the limitations of the chosen manufacturing method and two main design principles. DMLS is a flexible process, but is mainly limited by the necessity of support material and the size of the build platform. These limitations had to be considered throughout the design process. The first design principle that was implemented was designing the entire assembly to be produced on a standard DMLS machine. This limited the necessity to consider components that might complicate the design through interfacing between additive and traditionally machined components. The second design principle was ensuring that the designed load was broadband. This meant that there would be no resonant structures, which typically requires more precise control of the surface imperfections.

\par All simulations were performed within Ansys High-Frequency Structure Simulator (HFSS) software. Simulations were performed to observe the change each major component of the design (taper, spiral, pump-out hardware) had on the the overall performance. Initial designs were based off of those done in the X-band \cite{cern1} and modified to better match the frequency range of C-band. The goal of these models was to achieve a S$_{11}$ value of -20 dB or lower at 5.712 GHz with a wide bandwidth. The bandwidth was necessary to avoid problems due to manufacturing and tuning of the structure. The initial model generated was a straight line, non-standard waveguide without any taper to a standard WR-187 waveguide. This was done to ensure that the non-standard waveguide model allowed for broadband transmission of an arbitrary signal around the C-Band range.

\par
\begin{figure}[h]
\centering
\includegraphics[width=0.5\textwidth]{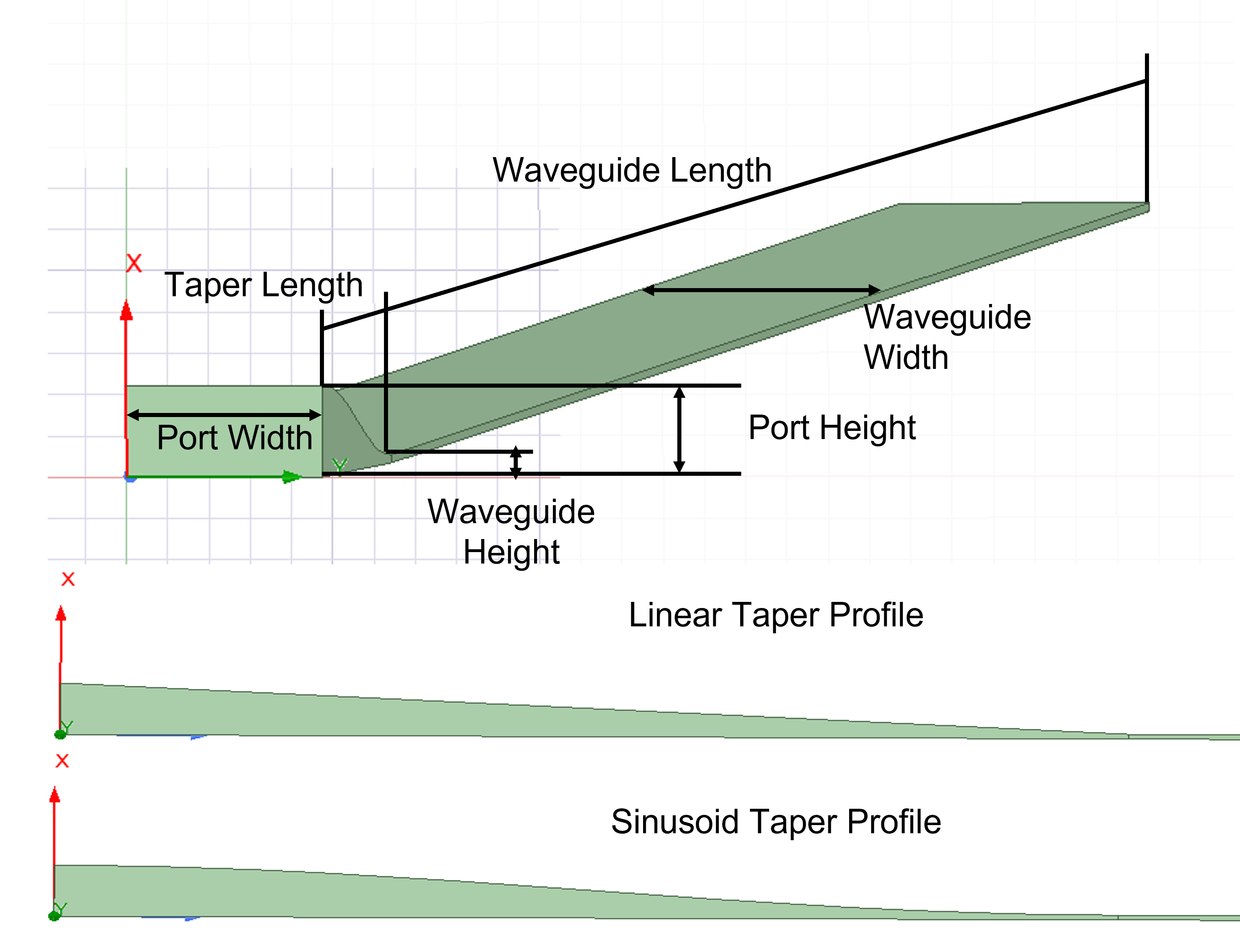}
\caption{Definition of dimensions shown in Table \ref{tab:Str8Def} along with the profiles of the different tapers.}\label{fig:Str8WGDim}
\end{figure}

 Tapers were used to transition from the standard WR-187 waveport to the non-standard waveguide. Two different tapers were tested in the height dimension, a linear taper, and a sinusoidal taper. A simple linear expansion from the narrower port width to the wider waveguide width was designed. Equation (\ref{eq:sinTaper}) defines the sinusoidal profile using the given variables from Table \ref{tab:Str8Def} to show both the linear and sinusoidal taper in Fig. \ref{fig:Str8WGDim}.
\captionsetup{labelsep=none}
\begin{table}[h]
\centering
\caption{\\Straight line model dimension definitions.}
\begin{tabular}{|c|c|c|}
    \hline
    Variable Name & Symbol & Value (mm)  \\
    \hline
    Port Width & $w_{wr}$ & 47.55 \\
    Port Height & $h_{wr}$ & 22.15\\
    Waveguide Width & $w_{wg}$ & 60\\
    Waveguide Height & $h_{wg}$ & 2\\
    Taper Length & $l_t$ & 400\\
    Waveguide Length & $l_{wg}$ & 7000\\
    \hline
\end{tabular}
\label{tab:Str8Def}
\end{table}
\captionsetup{labelsep=period}

\begin{equation}
x(z) = \frac{h_{wr}+h_{wg}}{2}+\frac{h_{wr}-h_{wg}}{2}\cos(\frac{\pi}{l_t}z)\;,\;z\in (0,l_t)
\label{eq:sinTaper}
\end{equation}

\par Once the models for the straight waveguide without any taper and the two with the linear and sinusoidal tapers were developed, models for the spiraled versions of them were generated. The straight portion of the waveguide was modelled by sweeping a rectangular face along a line defined by (\ref{eq:spiral}). Following the spiral equation of the straight waveguide, equations had to be developed for the two taper versions. The interior face of each taper was curved around the inner spirals along the line defined by (\ref{eq:spiral}) to match the straight models. The outer faces of the tapers had to be separately defined to maintain similar behavior as the straight line model. Equation (\ref{eq:spiralLinear}) shows the modified version of the linear taper in the spiral form and (\ref{eq:spiralSine}) show the modified version of  (\ref{eq:sinTaper}) in the spiral form.

\begin{equation}
\begin{cases}
    x(t)=(R+\alpha t)\sin(2\pi t)\\
    y(t)=(R+\alpha t)\cos(2\pi t)
\end{cases}
t\in (0,t_{l})
\label{eq:spiral}
\end{equation}

\begin{equation}
\alpha = h_{wg} + g
\end{equation}

\begin{equation}
\begin{cases}
    x(t)=(R+h_{wg} + (\alpha + \beta) t)\sin(2\pi t)\\
    y(t)=(R+h_{wg} + (\alpha + \beta) t)\cos(2\pi t)
\end{cases}
t \in (0,t_t)
\label{eq:spiralLinear}
\end{equation}

\begin{equation}
\beta = \frac{h_{wr}-h_{wg}}{t_t} 
\end{equation}

\begin{equation}
\begin{cases}
    x(t)=(R+h_{wg} + \kappa(t)\frac{t}{t_t})\sin(2\pi t)\\
    y(t)=(R+h_{wg} + \kappa(t)\frac{t}{t_t})\cos(2\pi t)
\end{cases}
t \in (0,t_t)
\label{eq:spiralSine}
\end{equation}

\begin{equation}\label{eq:endTapers}
        \kappa (t) = \left(\frac{h_{wg}-h_{wr}}{2}\right)\cos\left(\pi\frac{t}{t_t}\right) +\frac{h_{wg}}{4}+\frac{h_{wr}}{2}+gt_t
\end{equation}

\par Equations (\ref{eq:spiralLinear})-(\ref{eq:endTapers}) define the x and y coordinates that define the lines at $z(t)=0$, but must add (\ref{eq:widthTaper}) to add in the width taper to the sinusoidal and linear height taper

\begin{equation}
\label{eq:widthTaper}
z(t) = \frac{w_{wg}}{2} - \left(\frac{w_{wg}-w_{wr}}{2t_t}\right)t.
\end{equation}

\par
\begin{figure}[h]
\centering
\includegraphics[width=0.5\textwidth]{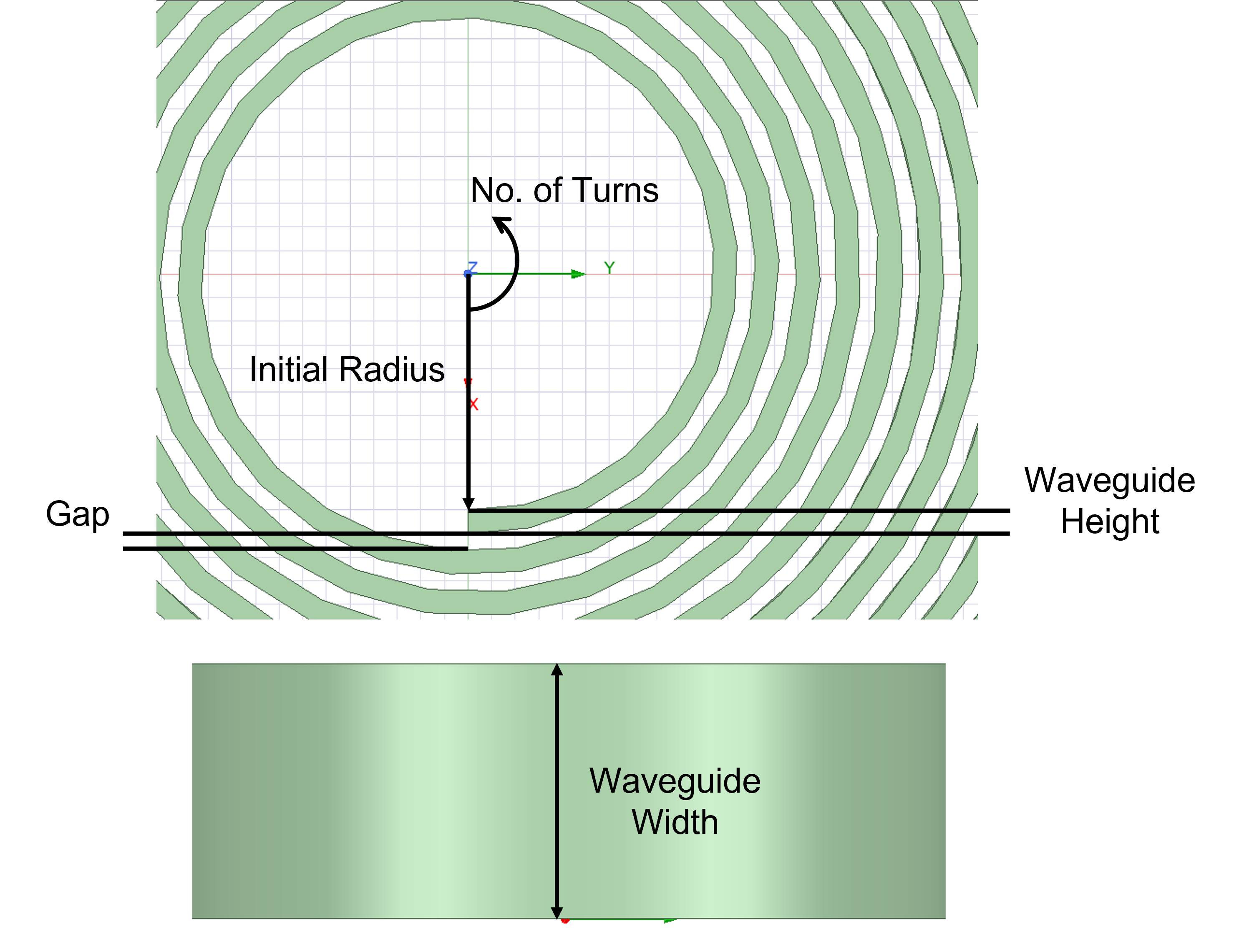}
\caption{Definition of dimensions shown in Table \ref{tab:SprDef}.}\label{fig:SprWGDim}
\end{figure}

 Fig. \ref{fig:SprWGDim} shows the way in which the various parameters of the spiral waveguide model was defined within HFSS. These values were modified slightly throughout the initial simulation runs to optimize the load performance. Table \ref{tab:SprDef}, along with $w_{wg}$ and $h_{wg}$ from Table \ref{tab:Str8Def}, define the parameters that were found to balance performance and print-ability of the spiral load structure.
\captionsetup{belowskip=0pt,aboveskip=0pt}

\begin{table}[h]
\captionsetup{labelsep=none}
\caption{\\Definitions of variables in Fig. \ref{fig:SprWGDim}.}
\centering
\begin{tabular}{|c|c|c|c|}
    \hline
     Variable Name & Symbol &  Value & Units\\
     \hline
     Gap & $g$ & 1.5 & mm\\
     Initial Radius & $R$ & 20 & mm  \\
     No. of Turns & $t_{l}$ & 19.5 & rev \\
     Taper Turn Length & $t_{t}$ & 0.75 & rev \\
     \hline
\end{tabular}
\label{tab:SprDef}
\end{table}

\par
\begin{figure}[h]
    \centering
    \includegraphics[width=0.5\textwidth]{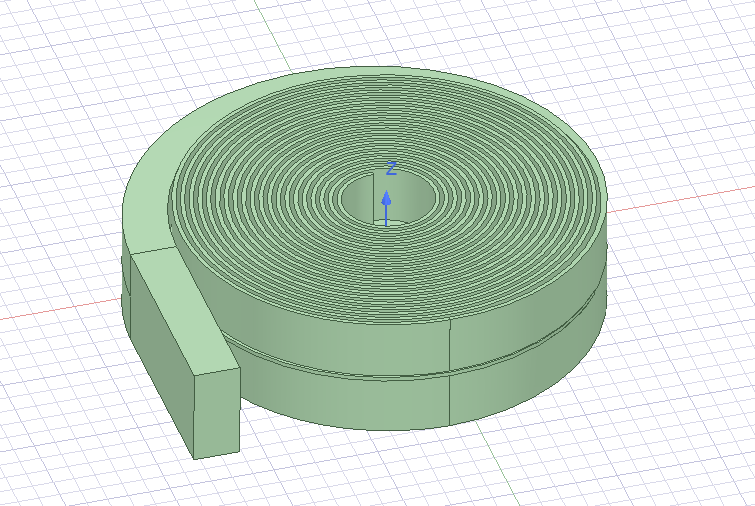}
    \caption{Final spiral load vacuum model and pump-out slits within the HFSS simulation software.}
    \label{fig:pumpoutSimImg}
\end{figure}

 The last development in the model was to add the pump-out hardware into the vacuum space model. This was an important step as it would allow the results of the simulation to more closely match the real-world performance of the load. It would also have to accommodate the need to remove powder from the interior of the load after the printing process. The model shown in Fig. \ref{fig:pumpoutSimImg} includes the pump out volume and WR-187 extension so that there was room for a flange to be added. The design shown also minimizes the inherent reflections caused by the slits and allows for the model to be split in two so that the final printed model is rendered in two halves. This allows for issues related to internal support material and powder removal to be mitigated.

\section{Results}
\subsection{Simulation Results}

\par
\begin{figure}
\centering
\includegraphics[width=0.5\textwidth]{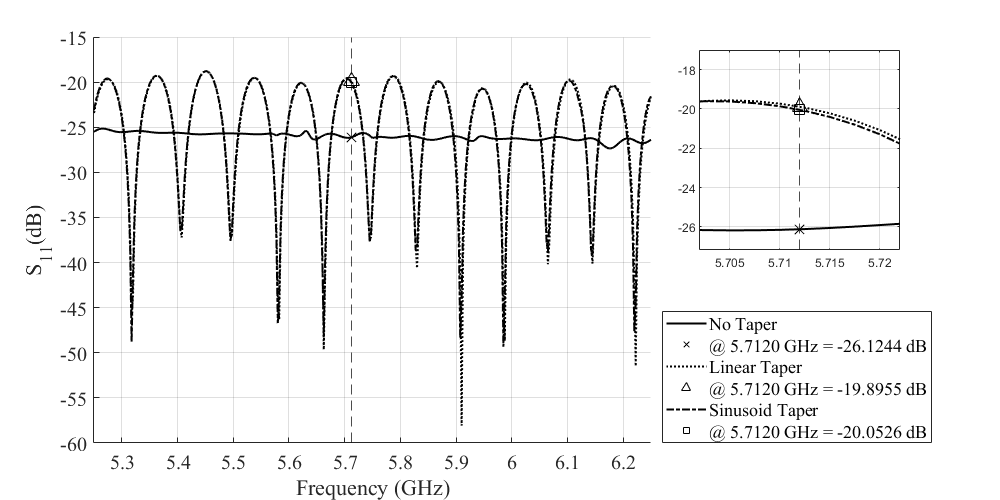}
\caption{Plot containing comparison of various taper methods on a straight waveguide model.}\label{fig:str8taperComp}
\end{figure}

 Using the straight waveguide model without any taper to a WR-187 port, it was then possible to compare the performance of the two different taper designs and dimensions shown in the Methodology section. Adding these tapers to the model and comparing it to the simulation run without the taper is shown in Fig. \ref{fig:str8taperComp}. This plot shows that there seems to be very little difference between the two taper designs in the straight load model. It shows slightly worse performance in the S$_{11}$ data when the taper is added when observing the changes at the desired center frequency of the load. This was expected to occur and can be explained by the additional reflections caused by the taper geometries. This is mainly due to the impedance mismatch from one side to another. It required further investigation into the potential effects that the tapers would have once the model was moved from a straight model to a spiral model to make a determination on what taper to utilize in the design.

\par
\begin{figure}
\centering
\includegraphics[width=0.5\textwidth]{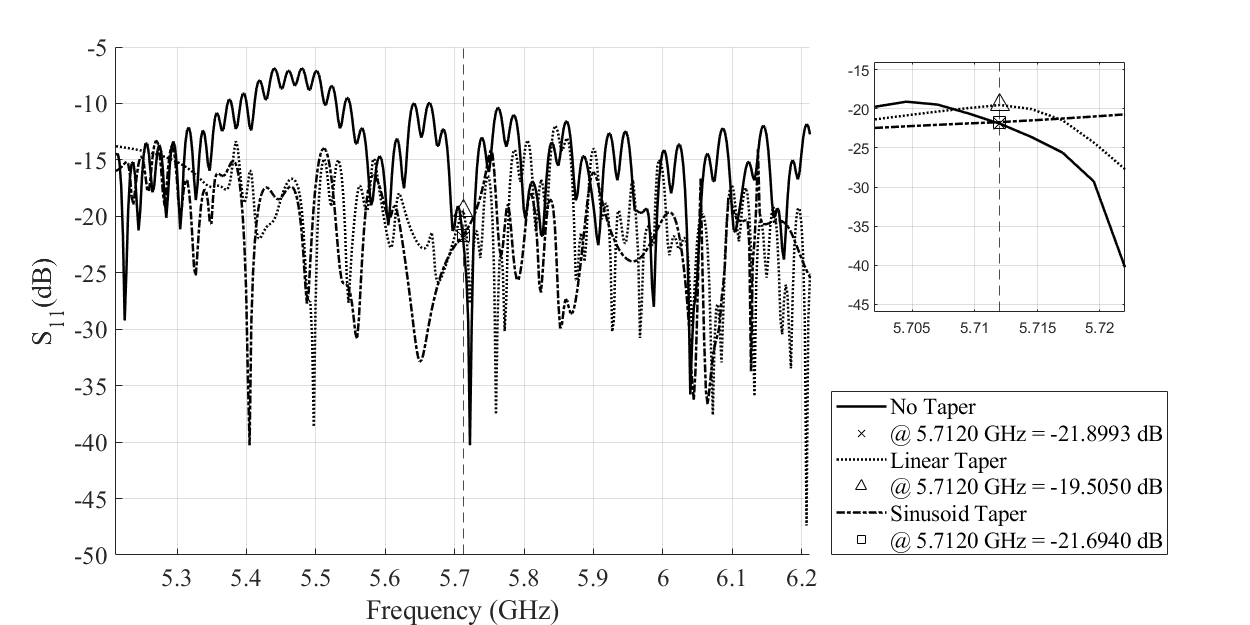}
\caption{Plot containing comparison of various taper methods on a spiral waveguide model.}\label{fig:spiralTrans}
\end{figure}

 Adding in the effect of the spiral load led to the ability to differentiate the effects of the sinusoidal taper and the linear taper. Fig. \ref{fig:spiralTrans} shows a plot of these differences. The sinusoidal taper produces an S$_{11}$ value of -22.7 dB at the desired frequency of 5.712 GHz. It performs better than the linear taper method (-19.5 dB) and similarly to the performance had there been no taper method utilized (-21.9 dB). For these reasons, it was decided to move on with the sinusoidal taper method in future simulation models as it showed better performance than that of the linear taper in the S$_{11}$ plot. 

\par
\begin{figure}[h]
\centering
\includegraphics[width=0.5\textwidth]{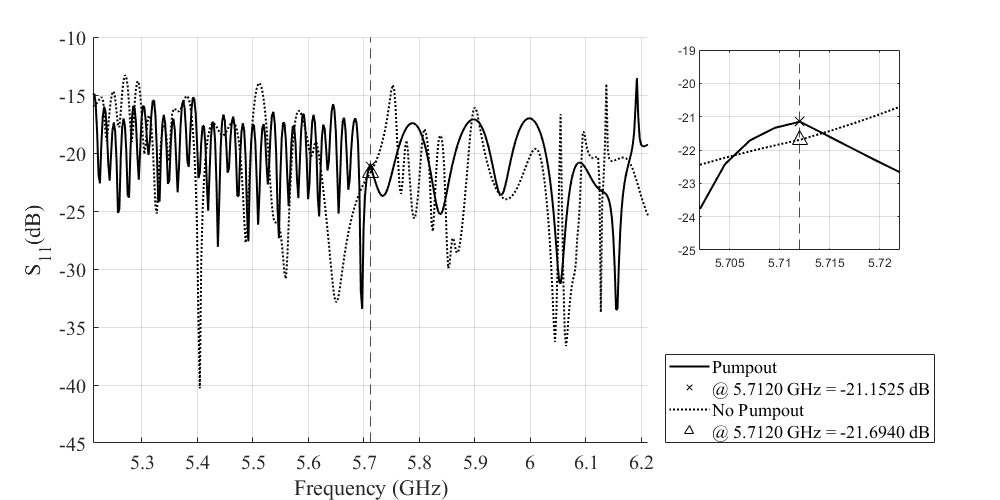}
\caption{Plot containing a comparison of the model with and without the pump-out volume.}\label{fig:pumpoutPlot}
\end{figure}

 Following the use of the spiral load model, the final pump-out model with all of the additional vacuum space designed for pump-out could be simulated. This was done to show the potential effects of the additional pump-out model might have on the reflections. This difference between the non-pump-out spiral model and model that added the pump-out hardware can be seen in Fig. \ref{fig:pumpoutPlot}. The plot shows a slight variation of the S$_{11}$ data across the frequency sweep. At the desired frequency of 5.712 GHz, the plots show a close match, with the pump-out model showing an S$_{11}$ value of -21.2 dB and the model without showing a response of -21.7 dB. This shows that the designed pump-out does not cause a significant negative impact to the performance of the spiral load. 

\subsection{Printed Model Results}

\par
\begin{figure}
    \centering
    \includegraphics[width=0.5\textwidth]{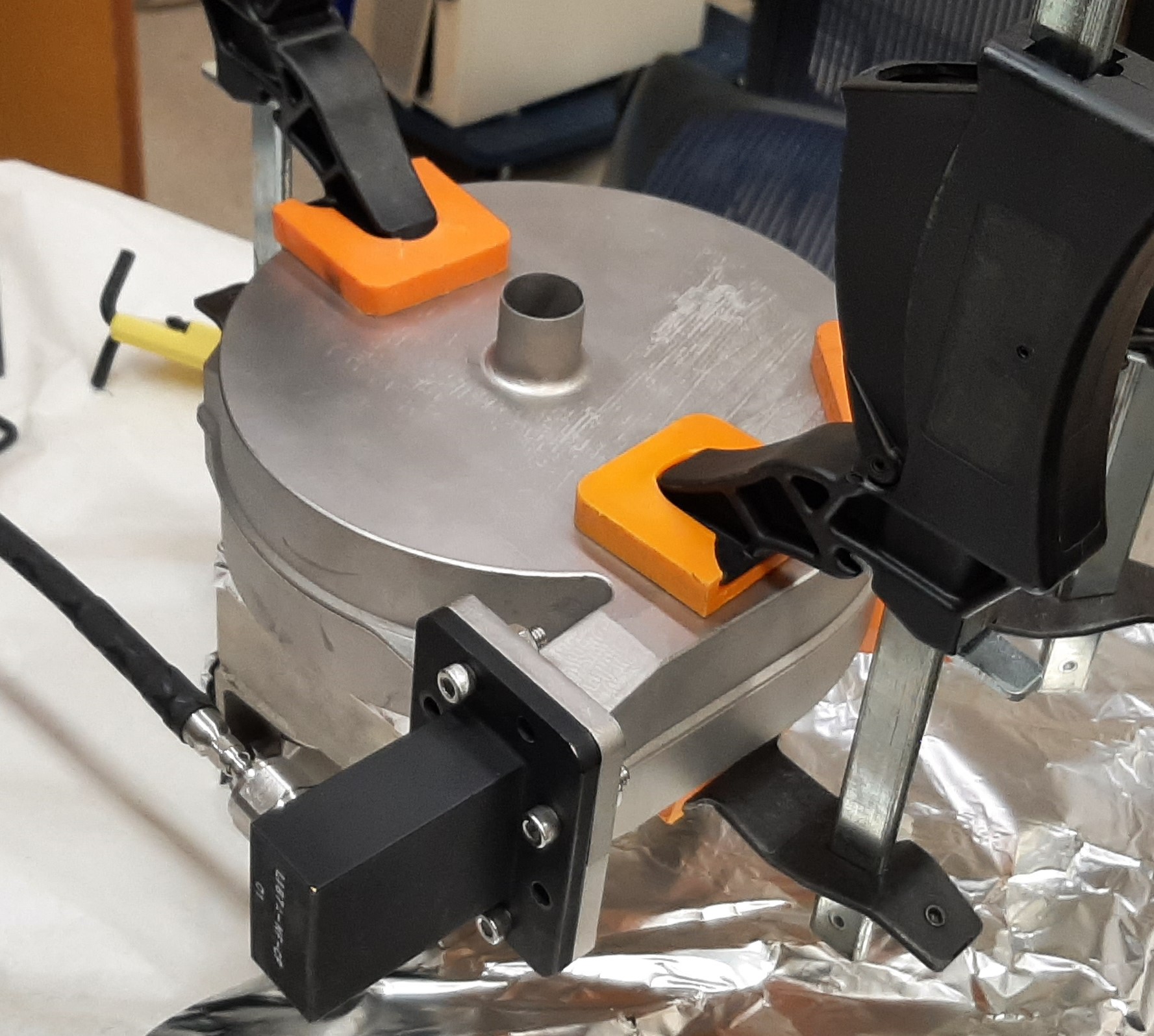}
    \caption{Photograph of initial cold test setup.}
    \label{fig:initColdTestimg}
\end{figure}

\begin{figure}
    \centering
    \includegraphics[width=0.5\textwidth]{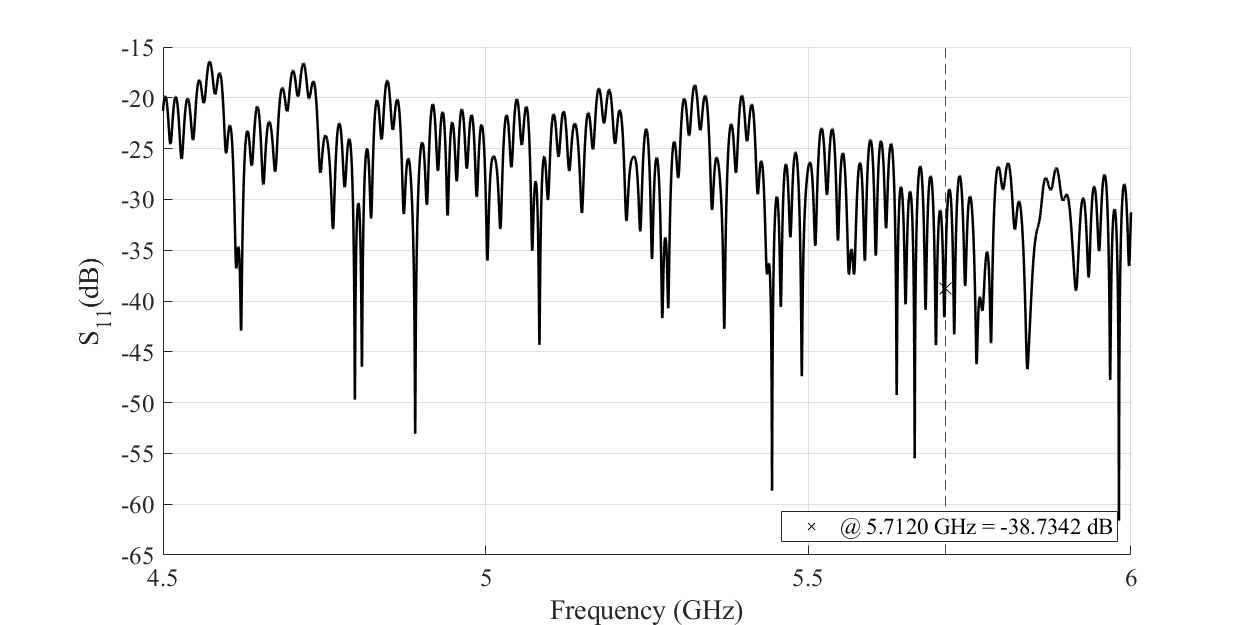}
    \caption{Plot of the S$_{11}$ curve of the spiral load during the initial cold test.}
    \label{fig:initColdTestplot}
\end{figure}

 Before the welding of the two halves of the spiral load, a cold test was performed for initial validations of the simulations and to verify that there were no unexpected results due to the printing process. The data for all of the cold tests was recorded on an Agilent N5241A vector network analyzer (VNA) which was calibrated using the TLR (thru, line, reflect) calibration method. The initial cold test was performed by inserting the alignment pins and clamping the two halves of the spiral load together. The spiral load was then connected to the VNA, a photograph of which can be seen in Fig. \ref{fig:initColdTestimg}. The results from this initial cold test, before any welding or machining occurred can be seen in Fig. \ref{fig:initColdTestplot}.

\par
\begin{figure} 
    \centering
    \includegraphics[width = 0.5\textwidth]{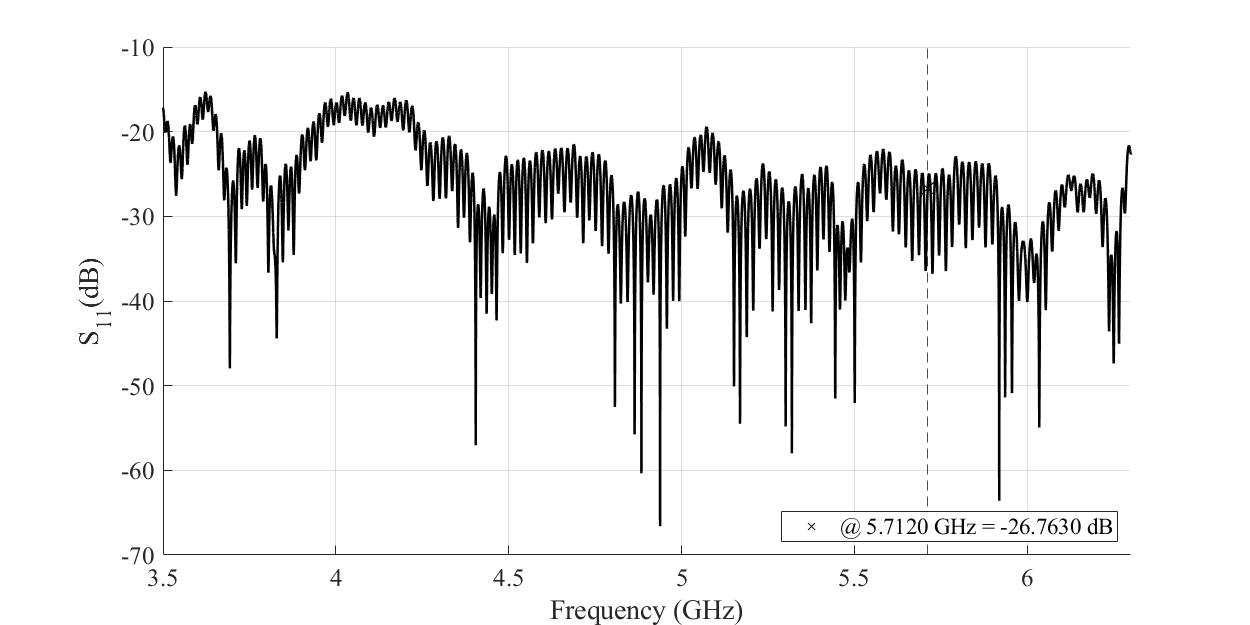}
    \caption{Plot of the S$_{11}$ curve of the spiral load following the welding of the two halves.}
    \label{fig:postWeldplot}
\end{figure}

 These results were promising, with the initial cold test showing that at a frequency of 5.712 GHz, the load was able to perform much better than expected according to simulations with an S$_{11}$ value of -38.7 dB. This result at the specific frequency is due to the fringing response being shifted and the overall power absorption is more consistent with the simulations. Following the initial cold test, the two halves were welded together and the instrumentation flange was kept to perform further tests on the load. Welding the two halves together had a significant effect on the S$_{11}$ curve, as shown in Fig. \ref{fig:postWeldplot}. This resulted in an S$_{11}$ value of -26.8 dB. While this was a significant difference at the specific frequency, the overall power absorption across the spectrum is consistent.

\par
\begin{figure}
    \centering
    \includegraphics[width=0.5\textwidth]{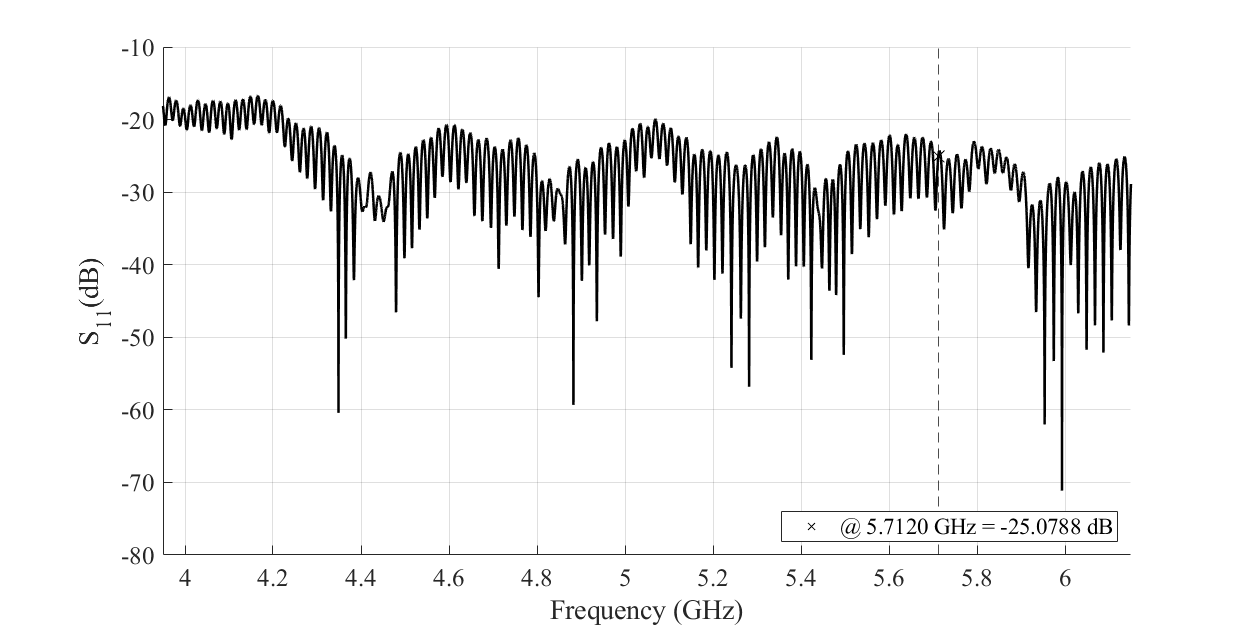}
    \caption{Plot of the S$_{11}$ response following the addition of the vacuum-rated hardware and cleaning of the load. This was recorded while the load was under atmospheric conditions.}
    \label{fig:flangeColdTest}
\end{figure}

\begin{figure}
    \centering
    \includegraphics[width = 0.5\textwidth]{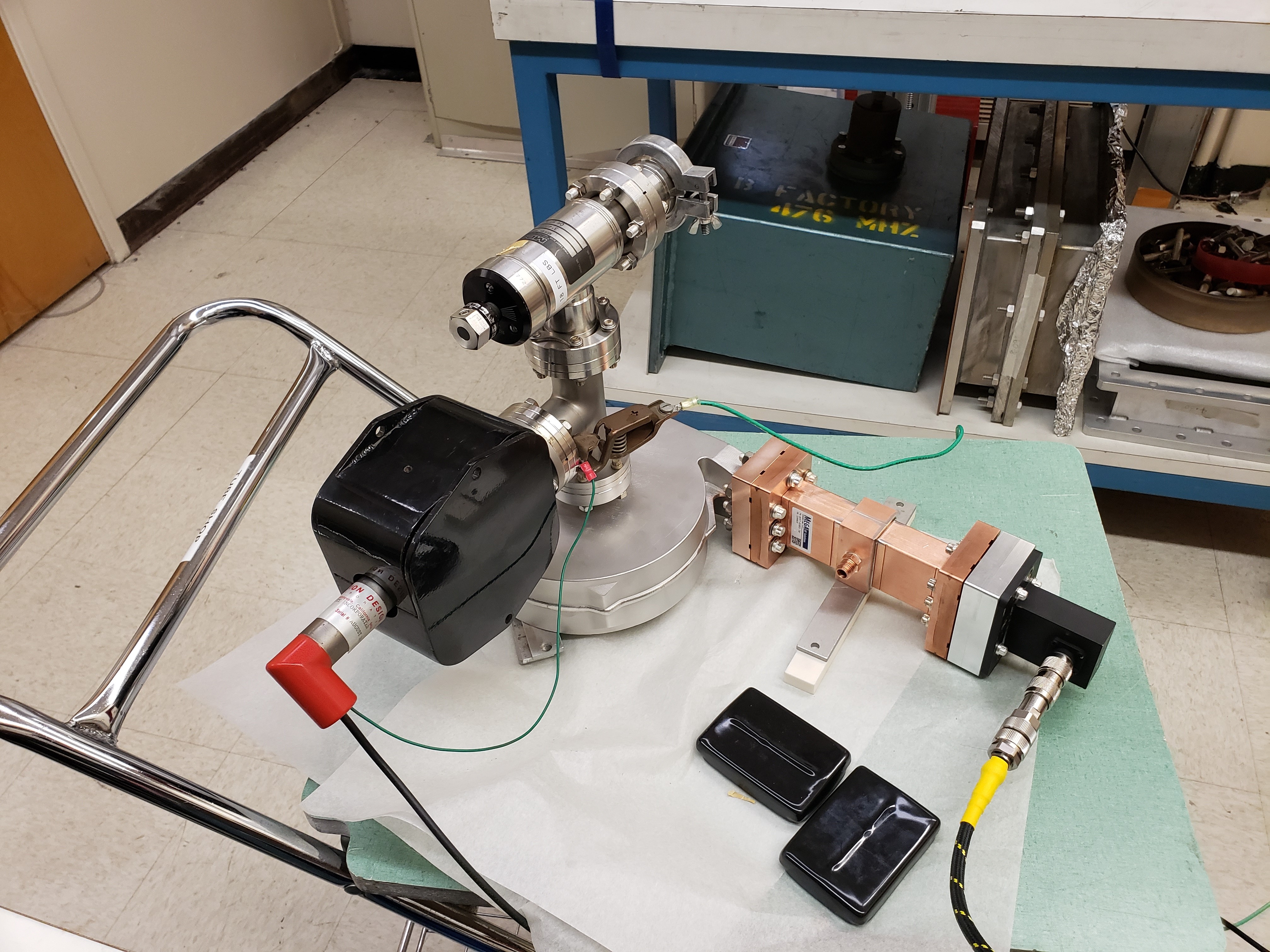}
    \caption{Image showing the bench setup of the spiral load when tested under vacuum conditions.}
    \label{fig:vacTestimg}
\end{figure}

\begin{figure}
    \centering
    \includegraphics[width = 0.5\textwidth]{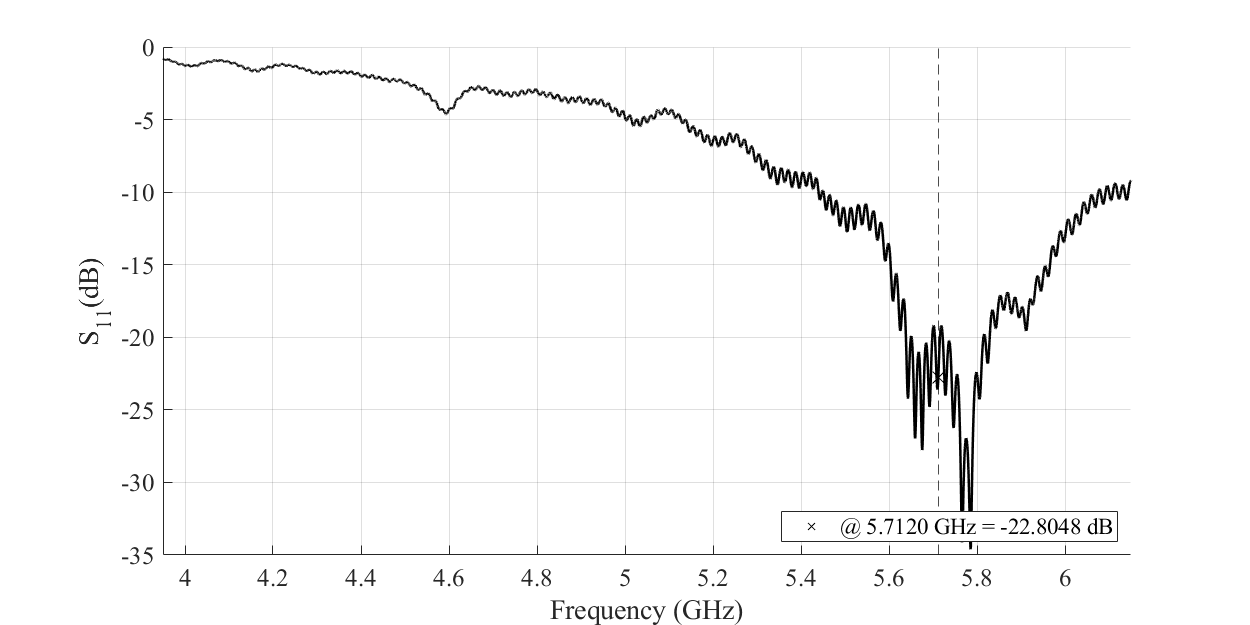}
    \caption{Plot of the S$_{11}$ curve during the cold test performed under vacuum conditions. The bandwidth is reduced due to the narrower bandwidth of the RF window.}
    \label{fig:vacTestplot}
\end{figure}

 So that the same printed load was used throughout all tests performed, the spiral load underwent the process of removing the instrumentation flange and high vacuum interfaces were added. Following this process, the spiral load was cold tested again under atmospheric conditions to see if there was any shift in the response due to additional welding and brazing processes. Fig. \ref{fig:flangeColdTest} shows the plot that resulted from this process under atmospheric conditions. It can be seen that at the center frequency of 5.712 GHz the S$_{11}$ value shifted up slightly from the previous cold test to a value of -25.1 dB. The load was then attached to a vacuum pump and an RF window was used to test the response while under vacuum conditions. A photograph of the bench setup is shown in Fig. \ref{fig:vacTestimg} and the results of the test are shown in Fig. \ref{fig:vacTestplot}. Adding the RF window and pumping the spiral load down to vacuum decreased the performance slightly to an S$_{11}$ value of -22.8 dB at the center frequency of 5.712 GHz. 

\par
\begin{figure}
    \centering
    \includegraphics[width=0.5\textwidth]{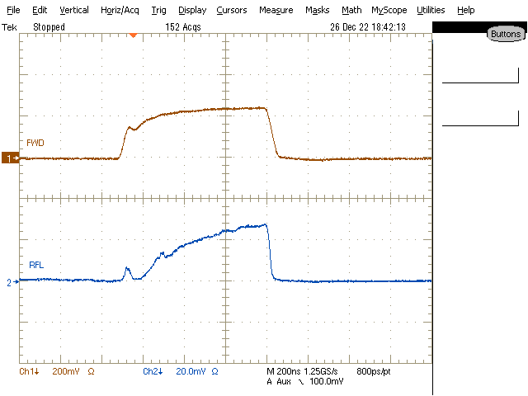}
    \caption{Screen capture of oscilloscope during testing with 8.1 MW power input at a 20 Hz rep rate and 700 ns pulse width.}
    \label{fig:8.1MW20Hz700ns}
\end{figure}

 The spiral load was sent to Radiabeam following all of the cold tests performed at SLAC. This was done to observe how the spiral load would perform while under high power loading. The load was first conditioned to prevent breakdown and damage to the spiral load structure. Conditioning began with a 200 ns pulse width at 47 kW of power, with a repetition rate of 1 Hz. The power and rep rate were then gradually stepped up to a rate of 20 Hz and maximum power of 8 MW. The pulse width was then increased once the load had been conditioned at the 200 ns pulse width to a pulse width of 400, 700, and finally 1000 ns. Once the load had been conditioned, the load was able to terminate a peak power of 8.1 MW with a repetition rate of 20Hz when the pulse width was set to 700 ns. Fig. \ref{fig:8.1MW20Hz700ns} shows the forward and reflected signal from the load during testing.

\par During the high power testing, the vacuum was able to stay stable at $3\times10^{-7}$ Torr when under a pulse width of 400~ns. The temperature of the spiral load, as measured by a thermocouple attached to the exterior of the load, showed a temperature increased from 22.8 \degree C to 27.9~\degree C over the testing period. When the load was tested with the 1000 ns pulse width, heating increased significantly. A peak temperature of 43.5 \degree C was present on the exterior of the load  and the vacuum increased from nominal of $3\times10^{-7}$ Torr to $1.3\times10^{-6}$ Torr causing testing at the 1000~ns pulse width to be paused. As the load was only passively cooled by the ambient atmosphere, a cooling fan was added and assisted in bringing down the peak temperatures to 34.9 \degree C and stabilized the pressure to $5.6\times10^{-7}$ Torr.

\section{Discussion}

\par The results show the feasibility of utilizing additive manufacturing to produce high-power C-band load structures. Using simulations in the initial development of the spiral load assisted in the ability to develop an operational solution. Two of the concerns in using simulations when compared to the manufactured model is the effect of slight misalignment ($<$0.1mm) of the two halves and the exact location and height of the surface roughness. The slight misalignment was due to slightly different shrinkage rates seen in the two halves leading to a discontinuity in the sidewalls. The surface roughness of the interior of the load is able to be approximately simulated but some of the regions of the load have a rougher surface than others, dependent on the angle that it was placed at within the DMLS machine. 

\par While some of these issues were present, it seems that they did not contribute to reduced performance and may have actually increased the performance of the load. This is mainly due to the additional resistance and surface area provided by the load due to the surface roughness. By utilizing additive manufacturing over traditional machining practices, the surface imperfections inherent in the process of manufacturing led to increased performance. As can be seen in the results, even after cleaning of the load, the S$_{11}$ at the design frequency is lower than the expected value based on the simulations performed.

\par While the load was able to terminate a peak power of 8.1 MW, testing on the load had to be stopped due to concerns over the stability of the vacuum pressure and the temperature on the external surface of the load. The increase in heat on the exterior following the finishing of testing suggested that there was a fairly significant amount of heat being produced on the interior walls of the spiral load, especially in the outer turns. This signifies that much of the power loss is concentrated in the first two or three turns of the spiral load. While is is known that the additional turns increased the overall performance, this concentration in surface loss led to uneven heating in the load and a reduction in the longevity of the performance of the load as vacuum pressure suffered. Future iterations of this design will need to include an integrated cooling system, an optimized waveguide shape to even out the surface loss concentrations, or both. These additions would allow for higher powers to be terminated, thus expanding the application of utilizing such processes.

\par The produced research shows the use of additive manufacturing techniques for the development of C-band load structures. The ability for additive manufacturing processes to adapt easily to design needs of the finalized structure allows greater flexibility in the manufacturing process. With the basic spiral load having proved the ability to perform at high powers allows for the advancement and further optimization for the specific application of the load. The results have shown promise in the future use of this technique in the C$^3$ concept. 

%\authorcontributions{Conceptualization, E.N., C.W., and J.M.; Methodology, C.W. and J.M.; Software, C.W.; Validation, G.M. and J.M.; Resources, G.M. and E.N.; Data Curation, B.S., J.M., R.A., A.D. and R.B.; Writing – Original Draft Preparation, G.M.  Writing – Review \& Editing, E.N., C.W., B.S., J.M. and G.M.; Visualization, G.M.; Supervision, E.N.;}

%\funding{This research was funded by the Department of Energy (DOE) under Contract No. DE-AC02-76SF00515.}

%\conflictsofinterest{The authors declare no conflicts of interest}

\bibliographystyle{IEEEtran}
\bibliography{references}

\begin{thebibliography}{10}
\providecommand{\url}[1]{#1}
\csname url@samestyle\endcsname
\providecommand{\newblock}{\relax}
\providecommand{\bibinfo}[2]{#2}
\providecommand{\BIBentrySTDinterwordspacing}{\spaceskip=0pt\relax}
\providecommand{\BIBentryALTinterwordstretchfactor}{4}
\providecommand{\BIBentryALTinterwordspacing}{\spaceskip=\fontdimen2\font plus
\BIBentryALTinterwordstretchfactor\fontdimen3\font minus
  \fontdimen4\font\relax}
\providecommand{\BIBforeignlanguage}[2]{{%
\expandafter\ifx\csname l@#1\endcsname\relax
\typeout{** WARNING: IEEEtran.bst: No hyphenation pattern has been}%
\typeout{** loaded for the language `#1'. Using the pattern for}%
\typeout{** the default language instead.}%
\else
\language=\csname l@#1\endcsname
\fi
#2}}
\providecommand{\BIBdecl}{\relax}
\BIBdecl

\bibitem{reviewOfAM}
\BIBentryALTinterwordspacing
T.~D. Ngo, A.~Kashani, G.~Imbalzano, K.~T. Nguyen, and D.~Hui, ``Additive
  manufacturing (3d printing): A review of materials, methods, applications and
  challenges,'' \emph{Composites Part B: Engineering}, vol. 143, pp. 172--196,
  2018. [Online]. Available:
  \url{https://www.sciencedirect.com/science/article/pii/S1359836817342944}
\BIBentrySTDinterwordspacing

\bibitem{terahertzAM}
B.~Zhang, Y.-X. Guo, H.~Zirath, and Y.~P. Zhang, ``Investigation on
  3-d-printing technologies for millimeter- wave and terahertz applications,''
  \emph{Proceedings of the IEEE}, vol. 105, no.~4, pp. 723--736, 2017.

\bibitem{amKlystron}
\BIBentryALTinterwordspacing
C.~Wehner, J.~Merrick, B.~Shirley, B.~Weatherford, G.~Mathesen, and E.~Nanni,
  ``Rf properties and their variations in a 3d printed klystron circuit and
  cavities,'' 2022. [Online]. Available:
  \url{https://doi.org/10.48550/arXiv.2211.15500}
\BIBentrySTDinterwordspacing

\bibitem{cern1}
\BIBentryALTinterwordspacing
H.~Bursali, ``X-band spiral load rf design,'' May 2022. [Online]. Available:
  \url{https://edms.cern.ch/document/2404437/1}
\BIBentrySTDinterwordspacing

\bibitem{industrialAdoption}
\BIBentryALTinterwordspacing
D.~Franco, G.~{Miller Devós Ganga}, L.~A. {de Santa-Eulalia}, and M.~{Godinho
  Filho}, ``Consolidated and inconclusive effects of additive manufacturing
  adoption: A systematic literature review,'' \emph{Computers \& Industrial
  Engineering}, vol. 148, p. 106713, 2020. [Online]. Available:
  \url{https://www.sciencedirect.com/science/article/pii/S0360835220304393}
\BIBentrySTDinterwordspacing

\bibitem{cern2}
\BIBentryALTinterwordspacing
H.~Bursali, ``Spiral loads rf measurement,'' May 2022. [Online]. Available:
  \url{https://edms.cern.ch/document/2323664/1}
\BIBentrySTDinterwordspacing

\bibitem{bai2021c}
M.~Bai, T.~Barklow, R.~Bartoldus, M.~Breidenbach, P.~Grenier, Z.~Huang,
  M.~Kagan, J.~Lewellen, Z.~Li, T.~W. Markiewicz \emph{et~al.}, ``C $^3$: A"
  cool" route to the higgs boson and beyond,'' \emph{arXiv preprint
  arXiv:2110.15800}, 2021.

\bibitem{dasu2022strategy}
S.~Dasu, E.~A. Nanni, M.~E. Peskin, C.~Vernieri, T.~Barklow, R.~Bartoldus,
  P.~C. Bhat, K.~Black, J.~Brau, M.~Breidenbach \emph{et~al.}, ``Strategy for
  understanding the higgs physics: the cool copper collider,'' \emph{arXiv
  preprint arXiv:2203.07646}, 2022.

\bibitem{nanni2022c}
E.~A. Nanni, M.~Breidenbach, C.~Vernieri, S.~Belomestnykh, P.~Bhat,
  S.~Nagaitsev, M.~Bai, W.~Berg, T.~Barklow, J.~Byrd \emph{et~al.}, ``C $^3$
  demonstration research and development plan,'' \emph{arXiv preprint
  arXiv:2203.09076}, 2022.

\bibitem{bane2018advanced}
K.~L. Bane, T.~L. Barklow, M.~Breidenbach, C.~P. Burkhart, E.~A. Fauve, A.~R.
  Gold, V.~Heloin, Z.~Li, E.~A. Nanni, M.~Nasr \emph{et~al.}, ``An advanced
  ncrf linac concept for a high energy e $^+ $ e $^-$ linear collider,''
  \emph{arXiv preprint arXiv:1807.10195}, 2018.

\end{thebibliography}

\end{document}